\documentclass[letterpaper,english,aps,prb,floatfix,showpacs,amsfonts,amssymb,superscriptaddress]{revtex4}
\usepackage[T1]{fontenc}
\usepackage[latin1]{inputenc}
\usepackage{graphics}
\usepackage{amssymb}
\usepackage{amsmath}
\usepackage{overpic}

\newcommand{\be}{\begin{equation}}
\newcommand{\ee}{\end{equation}}
\newcommand{\bea}{\begin{eqnarray}}
\newcommand{\eea}{\end{eqnarray}}

\newcommand{\rmi}{{\rm i}}

\newcommand{\cuoo}{CuO$_{2}$}
\newcommand{\lco}{La$_{2}$CuO$_{4}$}

\newcommand{\srcuocl}{Sr$_2$CuO$_2$Cl$_2$}

\def\d{\delta}

\def\m{\mu}

\def\D{\Delta}

\def\ra{\rightarrow}

\def\pll{\parallel}

\def\Ra{\Rightarrow}

\def\bk{{\bf k}}

\def\bQ{{\bf Q}}

\def\bH{{\bf H}}

\def\bL{{\bf L}}

\def\bS{{\bf S}}
\def\bn{{\bf n}}

\def\cA{{\cal A}}

\def\nn{\nonumber}
\def\lb{\label}
\def\pref#1{(\ref{#1})}

\newcount\bozza \bozza=0
\ifnum\bozza=1
\newdimen\shift \shift=-2truecm
\def\lb#1{%
{\label{#1}\rlap{\kern\shift{$\scriptstyle#1$}}}}
\else\def\lb#1{\label{#1}} \fi

\begin{document}

\title{Field dependence of the magnetic spectrum in anisotropic
  and Dzyaloshinskii-Moriya antiferromagnets: II. Raman
  spectroscopy}

\author{L.~Benfatto}

\email{lara.benfatto@roma1.infn.it}

\affiliation
{CNR-SMC-INFM and Department of Physics, University of Rome ``La
  Sapienza'',\\ Piazzale Aldo Moro 5, 00185, Rome, Italy}

\author{M.~B.~Silva~Neto}

\email{barbosa@phys.uu.nl}

\affiliation {Institute for Theoretical Physics, University of Utrecht,
  P.O. Box 80.195, 3508 TD, Utrecht, The Netherlands}

\author{A.~Gozar}

\email{agozar@bnl.gov}

\affiliation
{Bell Laboratories, Lucent Technologies, Murray Hill, NJ 07974}

\affiliation
{Brookhaven National Laboratory, Upton, New York 11973-5000}

\author{B.~S.~Dennis}

\affiliation
{Bell Laboratories, Lucent Technologies, Murray Hill, NJ 07974}

\author{G.~Blumberg}

\email{girsh@bell-labs.com}

\affiliation
{Bell Laboratories, Lucent Technologies, Murray Hill, NJ 07974}

\author{L.~L.~Miller}

\email{lancelmiller@gmail.com}

\affiliation
{Churchill H.S./Eugene SD 4J, Eugene, OR 97405 USA}

\author{Seiki~Komiya}

\email{komiya@komae.denken.or.jp}

\affiliation
{Central Research Institute of Electric Power Industry, 2-11-1
  Iwato-kita, Komae, Tokyo 201-8511, Japan}

\author{Yoichi~Ando}

\email{ando@criepi.denken.or.jp}

\affiliation
{Central Research Institute of Electric Power Industry, 2-11-1
  Iwato-kita, Komae, Tokyo 201-8511, Japan}

\date{\today}

\begin{abstract}

We compare the theoretical predictions of the previous article on the field
dependence of the magnetic spectrum in anisotropic two-dimensional and
Dzyaloshinskii-Moriya layered antiferromagnets [L.~Benfatto and M.~B.~Silva
Neto, cond-mat/0602419], with Raman spectroscopy experiments in {\srcuocl}
and untwinned {\lco} single crystals. We start by discussing the crystal
structure and constructing the magnetic point group for the magnetically
ordered phase of the two compounds, {\srcuocl} and {\lco}. We find that the
magnetic point group in the ordered phase is the $\underline{m}m\underline{m}$ 
orthorhombic group, in both cases. Furthermore, we classify all
the Raman active one-magnon excitations according to the irreducible 
co-representations for the associated magnetic point group. We find
that the in-plane (or Dzyaloshinskii-Moriya) mode belongs to the
$DA_g$ co-representation while the out-of-plane (XY) mode belongs to 
the $DB_g$ co-representation. We then measure and fully characterize
the evolution of the one-magnon Raman energies and intensities for
low and moderate magnetic fields along the three crystallographic
directions. In the case of {\lco}, a weak-ferromagnetic transition is
observed for a magnetic field perpendicular to the {\cuoo} layers. We
demonstrate that from the jump of the Dzyaloshinskii-Moriya gap at the
critical magnetic field $H_c\simeq 6.6$ T one can determine the value of
the interlayer coupling $J_\perp/J\simeq 3.2\times 10^{-5}$, in good
agreement with previous estimates. We furthermore determine the components
of the anisotropic gyromagnetic tensor as $g_s^a=2.0$, $g_s^b=2.08$, and
the upper bound $g_s^c=2.65$, also in very good agreement with earlier
estimates from magnetic susceptibility. For the case of {\srcuocl}, we
compare the Raman data obtained in an in-plane magnetic field with previous
magnon-gap measurements done by ESR. Using the very low magnon gap
estimated by ESR ($\sim 0.05$ meV), the data for the one-magnon Raman
energies agree reasonably well with the theoretical predictions for
the case of a transverse field (only hardening of the gap). On the
other hand, an independent fit of the Raman data provides an estimate
for $g_s\simeq 1.98$ and gives a value for the in-plane gap 
larger than the one measured by ESR. Finally, because of the absence
of the Dzyaloshinskii-Moriya interaction in {\srcuocl}, no
field-induced modes are observed for magnetic fields parallel to the 
{\cuoo} layers in the Raman geometries used, in contrast to the
situation in {\lco}.
\end{abstract}


\maketitle

\section{Introduction}

\begin{figure}[htb]
\includegraphics[scale=0.25]{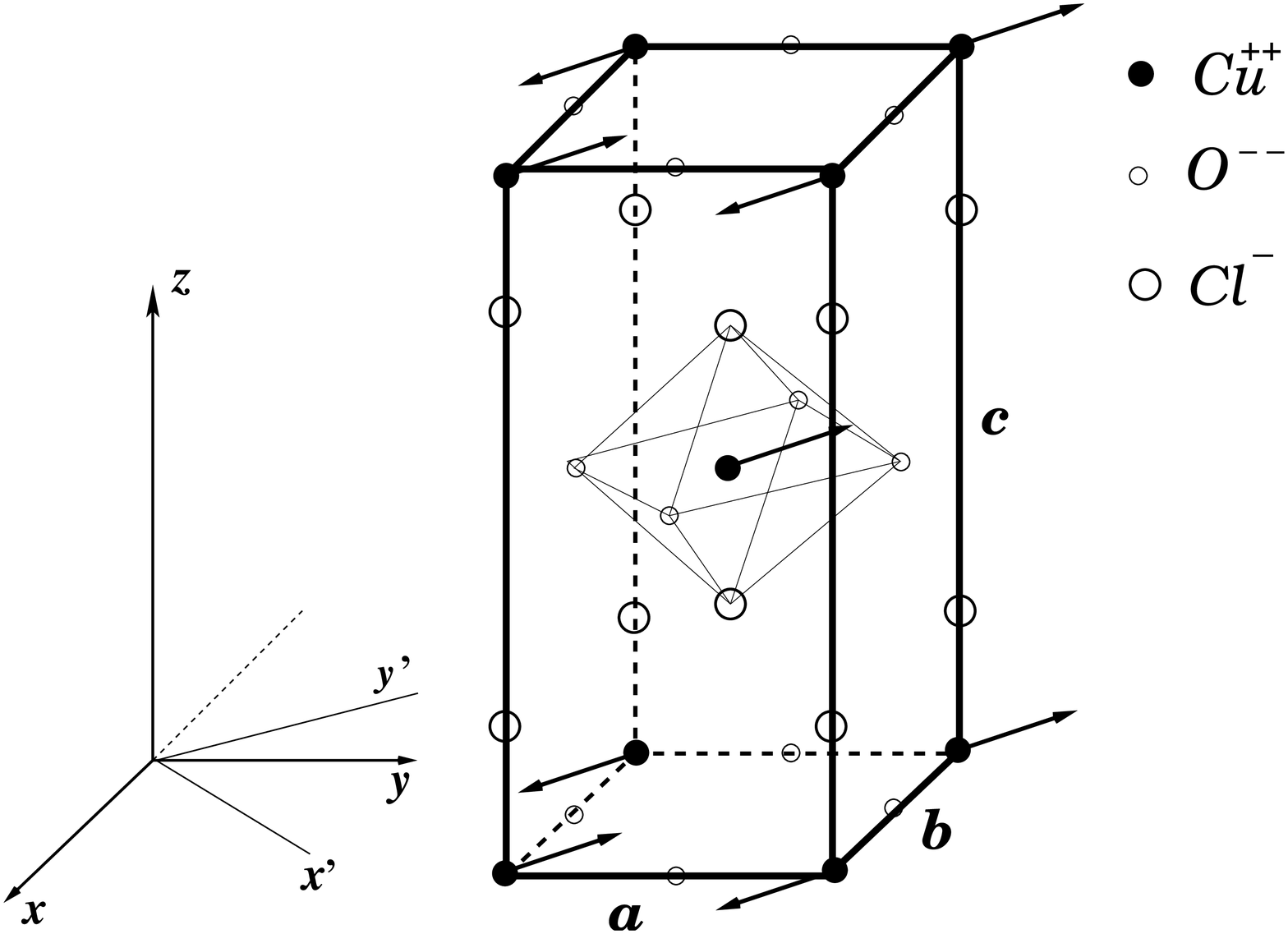}
\caption{Tetragonal magnetic unit cell of {\srcuocl} for $T<T_N$ (only
  Cu$^{++}$, O$^{--}$ and Cl$^{-}$ ions are shown for clarity). For the sake of
  unifying notations between {\srcuocl} and {\lco}, in what follows we
  shall use the $(abc)$ coordinate system above where $a\parallel x'$,
  $b\parallel y'$, and $c\parallel z$. Since the crystal structure is
  tetragonal, we have $a=b<c$. The Cu$^{++}$ spins are confined to the
  $(ab)$ plane (CuO$_2$ layers) and are oriented along the
  $(\overline{1}10)$ direction in the $(xyz)$ coordinate system, or,
  equivalently, parallel to the $b$ axis. In the paramagnetic phase, the
  point group is the $I4/mmm$ tetragonal group, but, as we shall discuss
  soon, below $T_N$ the symmetry is lowered because of the
  antiferromagnetic ordering of the spins.}
\label{Fig-SrCuOCl-Neel}
\end{figure}

In the preceding article,\cite{LM1} we investigated the field
dependence of the magnetic spectrum in anisotropic two-dimensional 
and Dzyaloshinskii-Moriya layered antiferromagnets. The first case 
is relevant to the understanding of the magnetic properties of 
{\srcuocl}, which is a body-centered tetragonal $S=1/2$ 
antiferromagnet\cite{Vaknin} with $I4/mmm$ structure\cite{Grande} and
$D_{4h}$ point group in the paramagnetic phase, $T>T_N$. Because of
its tetragonal character, at the classical level there should be no 
in-plane anisotropies present in {\srcuocl}. Furthermore, because of
its body-centered structure, $J_\perp$ is strongly frustrated.\cite{Greven} 
However, such perfect frustration can be removed by quantum fluctuations  
due to the spin-orbit interaction, and indeed a small in-plane
anisotropy is present in {\srcuocl},\cite{Yildirim} determining a spin
easy-axis at low temperatures, and giving rise to a very small
in-plane spin gap.\cite{ESR} For this reason, the magnetism in 
{\srcuocl} can be fairly well described by the following
two-dimensional square-lattice spin-Hamiltonian
\be
\lb{conv}
H_{con}=\sum_{\langle i,j\rangle}JS_i^bS_j^b+(J-\alpha_a)S_i^aS_j^a+
(J-\alpha_c)S_i^cS_j^c.
\ee
In the above expression, $S_i^{a,b,c}$ are the components along the
crystallographic axes $a,b,c$ (see Fig.\ \ref{Fig-SrCuOCl-Neel}) of the
Cu$^{++}$ spin, $J$ is the planar antiferromagnetic superexchange,
and $\alpha_a$ and $\alpha_c$ are, respectively, parameters that
control the in-plane and XY anisotropies. It should be emphasized 
here, once more, that at the classical level $\alpha_a=0$ because 
the crystal structure is body centered tetragonal.\cite{Kastner} A 
nonzero $\alpha_a$ can however be effectively obtained once quantum 
fluctuations (mostly from the spin-orbit coupling), which lift the 
frustration, are taken into account by considering the mean field 
effect of the neighboring layers.\cite{Greven}

%
%
%

\begin{figure}[htb]
\includegraphics[scale=0.24]{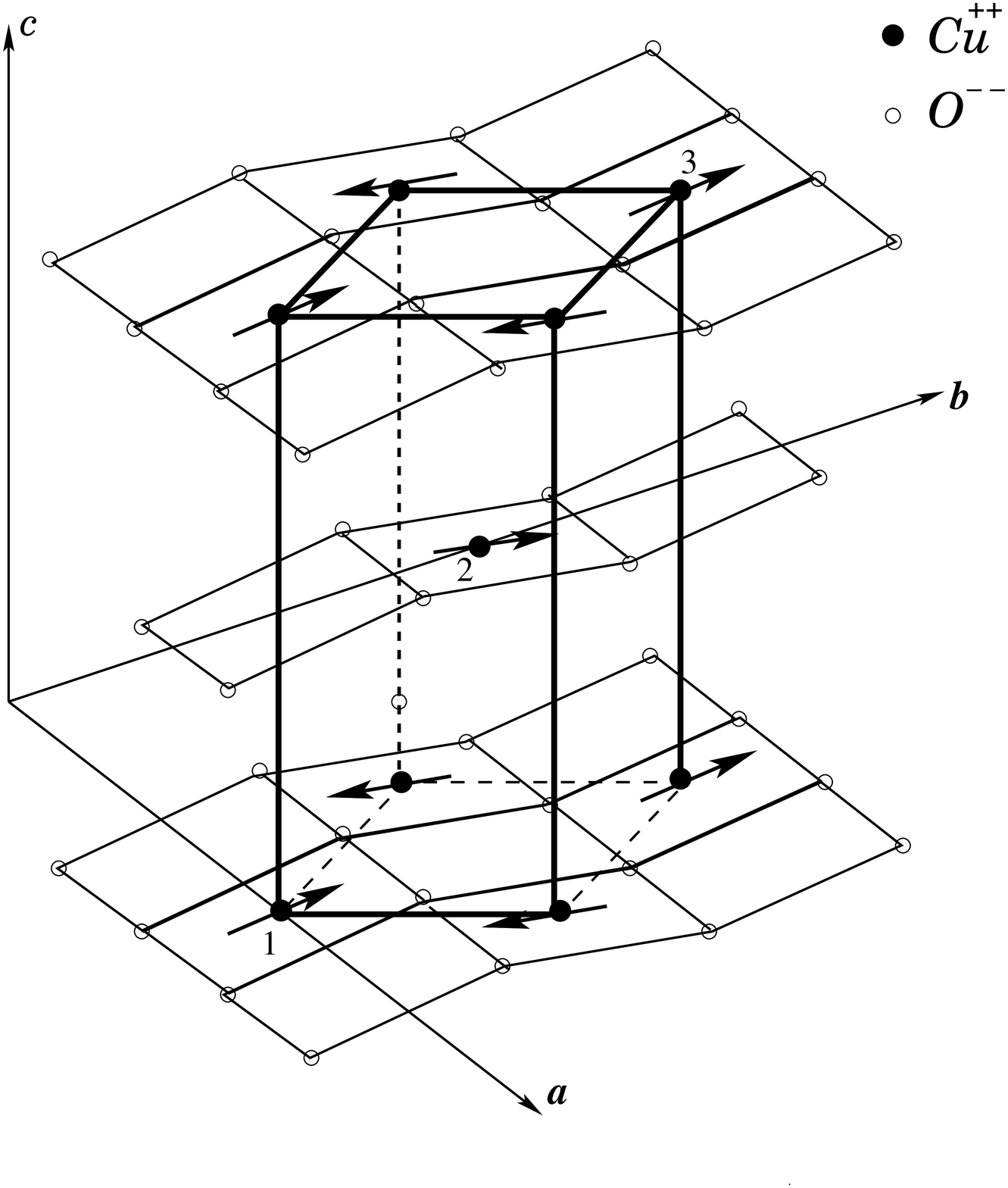}
\caption{(Color online) - Orthorhombic magnetic unit cell of {\lco} for
  $T<T_N$ (only Cu$^{++}$ and O$^{--}$ ions -except the apical ones- are
  shown for clarity). We use the $(abc)$ orthorhombic coordinate system,
  where $a\neq b\neq c$.  The Cu$^{++}$ ions order antiferromagnetically
  parallel to the $b$ orthorhombic axis but are also canted out of each
  CuO$_2$ layer, and are thus confined to the $(bc)$ plane. Observe that
  the canting pattern is staggered along the $c$ orthorhombic
  direction.}
\label{Fig-LCO-Neel-tilted}
\end{figure}

\begin{figure}[htb]
\includegraphics[scale=0.42]{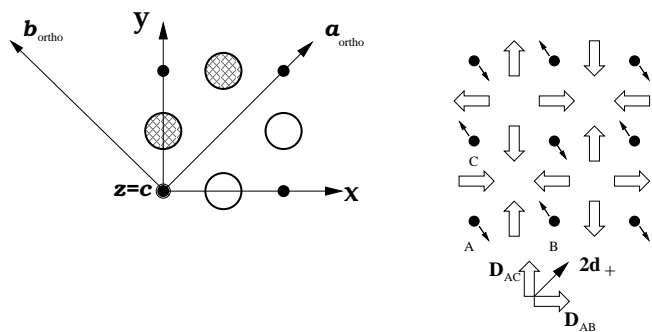}
\caption{Left: filled (small) circles represent the Cu$^{++}$ ions. The
  hatched larger circles represent the oxygen O$^{--}$ ions that are tilted
  above the CuO$_2$ layer while the open larger circles are O$^{--}$ ions
  that are tilted below the CuO$_2$ layer, see also Fig.\
  \ref{Fig-LCO-Neel-tilted}. Right: the staggered pattern of the
  Dzyaloshinskii-Moriya vectors, represented by open arrows along the Cu-Cu
  bonds, follows from the staggered pattern of the tilting of the oxygen
  octahedra. Small arrow: in-plane projection of the Cu$^{++}$ spins in
  the ordered phase.}
\label{Fig-DM-Vectors}
\end{figure}

Conversely, any realistic model for the magnetism of {\lco} that takes into
account the tilting of the oxygen octahedra should incorporate both a
Dzyaloshinskii-Moriya (DM) interaction and also the interlayer coupling
$J_\perp\neq 0$.\cite{Shekhtman} In the low-temperature orthorhombic
phase of {\lco}, $T<530$ K, the crystal has the $Bmab$ structure
with the $D_{2h}$ point group in the paramagnetic phase, $T>T_N$. The full
Hamiltonian that incorporates the Dzyaloshinskii-Moriya and XY interactions
allowed by symmetry, as well as the interlayer coupling, reads
\be
\lb{htot}
H=J_\perp\sum_m {\bf S}^m\cdot{\bf S}^{m+1}+\sum_m H_{sl}[{\bf S}^m,{\bf D}^m],
\ee
where ${\bf S}^m$ and ${\bf D}^m$ represent, respectively, the Cu$^{++}$ 
spins and Dzyaloshinskii-Moriya vectors at a generic lattice position
$(i,j)$ of the $m$th layer. The sum is over the Hamiltonian for a
single layer
\be
\lb{Hamiltonian}
H_{sl}[{\bf S},{\bf D}]=J\sum_{\langle i,j\rangle}{\bf S}_{i}\cdot{\bf S}_{j}+
\sum_{\langle i,j\rangle}{\bf D}_{ij}\cdot\left({\bf S}_{i}\times{\bf
    S}_{j}\right)+
\sum_{\langle i,j\rangle}{\bf S}_{i}
\cdot\overleftrightarrow{\bf \Gamma}_{ij}\cdot{\bf S}_{j},
\ee
where ${\bf D}_{ij}$ and $\overleftrightarrow{\bf \Gamma}_{ij}$ are,
respectively, the DM and XY anisotropic interaction terms that arise due to
the spin-orbit coupling and direct-exchange in the low-temperature
orthorhombic (LTO) phase of {\lco} (see Fig.\ \ref{Fig-DM-Vectors} and
the preceding article for a proper definition of these quantities). It
should be noted here that due to the peculiar staggered pattern of the
tilting angle of the oxygen octahedra in neighboring layers, the 
Dzyaloshinskii-Moriya vector alternates in sign from one layer to the other, 
${\bf D}^m_{AB,AC}=(-1)^m{\bf D}_{AB,AC}$. Moreover, since the unit cell is
body centered, the coupling $J_\perp$ in Eq.\ \pref{htot} connects the spin
at position (0,0,0) to the one at (1/2,0,1/2) in the LTO coordinate
system (see also Fig.\ \ref{Fig-LCO-Neel-tilted}).

As it has been discussed extensively in the preceeding article\cite{LM1},
the DM interaction leads to a quite unconventional response of the system
to an external magnetic field. Indeed, due to the DM interaction the spins
develop small out-of-plane ferromagnetic moments along with the staggered
ones characteristic of the AF order, which in turn couple to the external
field leading for example to unusual magnetic susceptibility anisotropies,
as measured at small fields in
{\lco}.\cite{Thio,Ando-Mag-Anisotropy,MLVC,Gooding} At larger field values
one can observe a ferromagnetic ordering of these moments along the $c$
axis for magnetic fields applied perpendicular to the {\cuoo}
layers,\cite{Thio,Thio90,Papanicolaou} or to a two-step spin-flop of the
staggered moments for an applied in-plane field
\cite{Thio90,Papanicolaou}. This physical picture, which was discussed in
the previous work of Refs.\
[\onlinecite{Shekhtman,Thio,Thio90,Papanicolaou}], has been confirmed by
the calculations reported in Ref.\ [\onlinecite{LM1}], where additional
results concerning the $(H,T)$ phase diagram and Raman response at finite
magnetic field have been discussed. This same approach turned out to be
quite successfull recently in demostrating that the Dzyaloshinskii-Moriya
interaction is behind the appearance of a field-induced mode in the
one-magnon Raman spectrum in {\lco} for longitudinal magnetic fields, as a
consequence of a rotation of the spin-quantization basis which was first
suggested in Ref. \onlinecite{Gozar}, theoretically explained in
Ref. \onlinecite{Marcello-Lara}, and quite recently observed in neutron
diffraction in Ref. \onlinecite{Reehuis}. Thus, the purpose of the present
paper is to directly compare the one-magnon spectrum measured in both
{\lco} and {\srcuocl} with the calculations presented in Ref.\
[\onlinecite{LM1}], as far as both the position and intensity of the Raman
peak is concerned. As we shall see, the excellent agreement between the
thoery and the experiments allows from one side to establish on firmer
grounds the general picture of the anomalous effect of the DM interaction in
the {\lco} system, and from the other side to estimate some relevant
physical quantities as the spin gyromagnetic ratio.

%
%
%

%
%
%

\section{Magnetic Group Analysis for $\mbox{\lco}$ and $\mbox{\srcuocl}$}

Before introducing the experimental setup, a preliminary discussion is
needed about the point group of $\mbox{\lco}$ and $\mbox{\srcuocl}$ both in
the paramagnetic and antiferromagnetic phases, in order to establish the
irreducible co-representations of the Raman tensors. While the paramagnetic
case has been already extensively discussed in the literature (see for
example Ref. \onlinecite{Reehuis} and references therein), a discussion of
the magnetic phase is still missing, and we shall briefly review here the
basic steps needed to make this analysis.

Let us consider a crystal with a
certain symmetry group, ${\bf G}$, and its set of allowed symmetry group
operations in the paramagnetic phase, $T>T_N$. When the system orders
antiferromagnetically, $T<T_N$, usually the symmetry is lowered because now
not all the sites are equivalent, and the corresponding magnetic group must
be determined.\cite{Birss} In general, two situations can occur:

\begin{itemize}

\item[(a)] Let us call ${\bf H}$ the restricted unitary subgroup of ${\bf
  G}$ containing the symmetry operations still allowed below $T_N$.
If ${\bf H}$ is a subgroup of ${\bf G}$ of index 2 (i.e. ${\bf H}$
contains exactly $n/2$ elements of ${\bf G}$) then {\em all} the
remaining elements of ${\bf G}-{\bf H}$ can nevertheless be promoted to
allowed symmetry operations when combined with the time-reversal operation
$\theta$, and are thus called anti-unitary elements. One can then identify,
for $T<T_N$, the magnetic point group corresponding to the classical point
group ${\bf G}$ as
\be
\lb{index2}
{\bf M}={\bf H}+\theta({\bf G}-{\bf H}).
\ee

\item[(b)]
However, it is possible that below $T_N$ the unitary operations still
allowed form a subgroup ${\bf \cal H}$ of index larger than 2 for ${\bf
G}$, but corresponding to a subgroup of index 2 of a different classical
group ${\bf \cal G}$ (i.e. ${\bf \cal H}$ contains $r/2<n/2$ elements,
where $r$ is the dimension of the group ${\bf \cal G}$). In this case the
magnetic group is identified by ${\bf \cal H}$ and the anti-unitary group
$\theta({\bf \cal G-H})$, so that
\be
\lb{index<2}
{\bf M}={{\bf \cal H}+\theta({\bf\cal  G}-{\bf \cal H})}.
\ee
\end{itemize}

\subsection{Magnetic point group for {\lco}}

The crystal structure of the low temperature orthorhombic phase
of {\lco} is the $Bmab$ structure in Fig.\ \ref{Fig-LCO-Neel-tilted}, 
which has as unitary group the $D_{2h}$ point group in the paramagnetic 
phase $T>T_N$. Above the N\'eel ordering temperature the allowed symmetry 
operations of the $D_{2h}$ group are
\be
D_{2h} = 1,{\;} \bar{1},{\;} 2_a,{\;} 2_b,{\;}
2_c,{\;} \bar{2}_a,{\;} \bar{2}_b,{\;} \bar{2}_c,
\ee
where the $8$ elements have their usual meanings\cite{Birss}: 1= identity;
$\bar 1$= inversion through the symmetry center, which is in this case the
central Cu$^{++}$ ion in position 2 of Fig.\ \ref{Fig-LCO-Neel-tilted}, so
that $(a,b,c) \rightarrow (-a,-b,-c)$; $n_z$= rotation of $2\pi/n$ around
the $z$ axis; $\bar n_z$= inversion through the symmetry center followed by
a $2\pi/n$ rotation around the $z$ axis (note that $\bar 2_z$ corresponds
also to a reflexion with a mirror perpendicular to the $z$ axis).

Although it may seem at first that neither $2_b$ nor $2_c$ are symmetry
operations, in both cases the final atomic configurations can be brought
back to the original one by a translation of half of the diagonal along
the $(011)$ direction (in the $(abc)$ coordinate system). In this sense, 
the octahedra in position $1$ (front-bottom-left corner) is brought to position 
$2$ (central ion) and the one from position $2$ is brought to position
$3$ (back-top-right corner). Such translation is allowed because there is no 
way to distinguish the corner and central Cu$^{++}$ ions in the crystal.

Below $T_N$, the Cu$^{++}$ ions order antiferromagnetically in the
pattern shown in Fig.\ \ref{Fig-LCO-Neel-tilted}. We can verify that
the remaining allowed {\it unitary} symmetry operations are
$$
{\bf H}= 1,{\;} \bar{1},{\;} 2_b,{\;} \bar{2}_b,
$$
bearing in mind that a half translation along the $(011)$ diagonal is
allowed. We immediately conclude that {\lco} belongs to the case (a)
discussed above, where the subgroup of allowed unitary operations is of
index $2$, see Eq.\ \pref{index2}. We can now construct the magnetic group
of the ordered phase of {\lco} by combining all the ${\bf G}-{\bf H}$ {\it
unitary} operations that are not allowed below $T_N$ with the time reversal
operation $\theta$ (represented in what follows by an underline), thus
forming {\it anti-unitary} operations. The allowed {\it unitary} $+$ {\it
anti-unitary} operations for {\lco} below $T_N$ are
\be
\lb{mglco}
{\bf M}={\bf H}+\theta({\bf G}-{\bf H})= 
1,{\;} \bar{1},{\;} \underline{2}_a,{\;} 2_b,{\;}
\underline{2}_c,{\;} \bar{\underline{2}}_a,{\;} \bar{2}_b,{\;}
\bar{\underline{2}}_c,
\ee
such that the magnetic point group for {\lco} below $T_N$ is the
$$
{\bf M}=\underline{m}m\underline{m}
$$
orthorhombic group, which also has $8$ elements. The Raman tensors for such
magnetic group are given in terms of the co-representations (see
Cracknell\cite{Cracknell})
\bea
\lb{Corep-ortho-mag}
DA_g{=}
  \left( \begin{array}{ccc}
     A & \rmi B & 0 \\
     \rmi D & E & 0 \\
     0 & 0 & I
         \end{array} \right),{\;}\quad
DB_g{=}
  \left( \begin{array}{ccc}
     0 & 0 & C \\
     0 & 0 & \rmi F \\
     G & \rmi H & 0
         \end{array} \right),{\;}
\eea
where $A,B,C,D,E,F,G,H,I$ are unconstrained real numbers.

\subsection{Magnetic point group for {\srcuocl}}

The crystal structure of the high temperature tetragonal (HTT) phase
of {\srcuocl} is the $I4/mmm$ structure in Fig.\ \ref{Fig-SrCuOCl-Neel}, 
which has as unitary group the $D_{4h}$ point group in the 
paramagnetic phase $T>T_N$. Above the N\'eel ordering temperature the 16 
allowed symmetry operations of the $D_{4h}$ group are\cite{Birss}
$$
D_{4h} = 1,{\;} \bar{1},{\;} 2_x,{\;} 2_y,{\;}
2_z,{\;} 2_{x^\prime},{\;} 2_{y^\prime},{\;} \bar{2}_x,{\;}
\bar{2}_y,{\;} \bar{2}_z,{\;} \bar{2}_{x^\prime},{\;}
\bar{2}_{y^\prime},{\;}\pm 4_z,{\;}\pm \bar{4}_z,
$$
where we used the reference system of Fig.\ \ref{Fig-SrCuOCl-Neel}.
As we can see, the tetragonal character of the unit cell allows for a
$4$-fold axis, the $c$-axis, which is perpendicular to the {\cuoo} planes
(see Fig.\ \ref{Fig-SrCuOCl-Neel}).

In the antiferromagnetic phase the spin easy-axis is along the $y^\prime$ or
$(010)$ direction in the $(abc)$ coordinate system, see Fig.\
\ref{Fig-SrCuOCl-Neel}. Thus, one can easily verify that, below $T_N$,
$2_x$, $2_y$, and $\pm 4_z$ are no longer symmetry operations, not even when
supplemented by the time reversal operation $\theta$. The only
allowed unitary operations in the N\'eel ordered phase of {\srcuocl}
are
$$
{\cal H}= 1,{\;} \bar{1},{\;} 2_{y^\prime},{\;} \bar{2}_{y^\prime},
$$
again bearing in mind that a half translation along the $(011)$ diagonal is
allowed. We immediately conclude that {\srcuocl} belongs to the case (b)
discussed above, as in Eq. \pref{index<2}, where the subgroup $\cal H$
of allowed unitary operations is a subgroup of $D_{4h}$ with index
larger than $2$, but it is a subgroup of ${\cal G}=D_{2h}$ ($mmm$) with index
$2$. We can now construct the magnetic group of the ordered phase of
{\srcuocl} by combining all the ${\cal G}-{\cal H}$ {\it unitary }
operations that are not allowed below $T_N$ with the time reversal 
operation $\theta$, thus forming {\it anti-unitary} operations. The 
allowed {\it unitary} $+$ {\it anti-unitary} operations for {\srcuocl} 
below $T_N$ are
\be
\lb{mgscoc}
{\bf M}={{\bf \cal H}+\theta({\bf\cal  G}-{\bf \cal H})}= 
1,{\;} \bar{1},{\;} \underline{2}_{x^\prime},{\;} 2_{y^\prime},{\;}
\underline{2}_z,{\;} \bar{\underline{2}}_{x^\prime},{\;} \bar{2}_{y^\prime},{\;}
\bar{\underline{2}}_z,
\ee
and thus the magnetic group, with only $8$ elements, is again the
orthorhombic $\underline{m}m\underline{m}$ group, just like the case of
{\lco}.\cite{Discussion-NiF2} In fact, when expressed in terms of the
$(abc)$ coordinate system of Fig.\ \ref{Fig-SrCuOCl-Neel}, such that
$a\parallel x'$, $b\parallel y'$, and $c\parallel z$, Eq.\ \pref{mgscoc}
can be written:
$$
{\bf M}= 1,{\;} \bar{1},{\;} \underline{2}_{a},{\;} 2_{b},{\;}
\underline{2}_c,{\;} \bar{\underline{2}}_{a},{\;} \bar{2}_{b},{\;}
\bar{\underline{2}}_c,
$$
which coincides with Eq.\ \pref{mglco} above. Thus, to unify the notation
we shall use in the following the $(abc)$ coordinate system for both {\lco}
and {\srcuocl} while discussing the antiferromagnetic phase. Moreover,
since the magnetic group is the same, the Raman tensors are given by Eq.\
\pref{Corep-ortho-mag} for both systems.

\section{Inelastic light scattering by magnons in
  $\mbox{La$_2$CuO$_4$}$ and $\mbox{\srcuocl}$}

One of the possible mechanisms for the inelastic scattering of light
by magnetic excitations in crystals is an indirect electric-dipole (ED) 
coupling via the spin-orbit interaction.\cite{Fleury-Loudon} Such a
mechanism has been in fact used to determine the spectrum of magnetic
excitations in many different condensed matter systems like fluorides,
XF$_{2}$, where X is Mn$^{2+}$, Fe$^{2+}$, or Co$^{2+}$,\cite{Fleury-Loudon} 
inorganic spin-Peirls compounds, CuGeO$_{3}$,\cite{Spin-Pierls} and the 
parent compounds of the high-temperature superconductors.\cite{Gozar}
The Hamiltonian representing the interaction of light with
magnons can be written quite generally as\cite{Cottam-Lockwood}
$$
H_{ED}=\sum_{{\bf r}}{\bf E}_S^T\chi({\bf r}){\bf E}_I,
$$
where ${\bf E}_S$ and ${\bf E}_I$ are the electric fields of the
scattered and incident radiation, respectively (${\bf a}^T$ is the
transposed of the ${\bf a}$ vector) and $\chi({\bf r})$ 
is the spin dependent susceptibility tensor. We can expand  
$\chi({\bf r})$ in powers of the spin-operators, ${\bf S}$, as
$$
\chi^{\alpha\beta}({\bf r})=\chi^{\alpha\beta}_0({\bf r})+
\sum_\mu K_{\alpha\beta\mu}S^\mu({\bf r})
+\sum_{\mu\nu} G_{\alpha\beta\mu\nu}
S^\mu({\bf r}) S^\nu({\bf r})
+\dots,
$$
where $\mu,\nu=x,y,z$ label the spin components. The lowest order 
term $\chi^{\alpha\beta}_0({\bf r})$ is just the susceptibility 
in the absence of any magnetic excitation (it corresponds to elastic 
scattering), and it will be neglected in what follows. The second and 
third terms can give rise to one-magnon excitations because they can 
be written as $S^\pm({\bf r})$ and $S^z({\bf r})S^\pm({\bf r})$,
respectively, where here $z$ is the direction of the spin easy-axis. 
The intensity of the scattering, as well as the selection rules, will
be determined by the structure and symmetry properties of the complex
tensors $K$ and $G$. 

The ED Hamiltonian that describes the ${\bf k}=0$ one-magnon
absorption/emission on sublattice A/B can now be written in terms
of the $x,y,z$ components of the sublattice magnetization, 
${\bf M}_i=({\bf S}_{i_A}-{\bf S}_{i_B})/2$, as
\be
H_{ED}=\sum_{i}\left\{ {\bf E}_S^T \chi^x {\bf E}_I{\;}M^x_i
+ {\bf E}_S^T \chi^y {\bf E}_I{\;}M^y_i\right\},
\label{H-ED}
\ee
where the matrices $\chi^x$ and $\chi^y$ are both written in terms of
the original $K$ and $G$ tensors. Here we made the usual mean-field
assumption $\langle S^z_{i_A}\rangle=-\langle S^z_{i_B}
\rangle=-S$ and we dropped terms of the type $S_{i_A}^{x,y}+S_{i_B}^{x,y}$ 
since these do not contribute for the ${\bf k}=0$ scattering. 

\subsection{Magnetic excitations and selection rules for 
  $\mbox{La$_2$CuO$_4$}$ and $\mbox{\srcuocl}$}

It is very important to emphasize here that, differently from two-magnon
Raman scattering, where the Raman response does not depend on the direction
of the spin easy-axis, the one-magnon Raman response does. Indeed, it has
been shown by Fleury and Loudon\cite{Fleury-Loudon} that one of the
incoming/outgoing components of the electric field must always lie in the
direction of the easy axis, and the other one in the perpendicular plane,
oriented in the direction of the mode that one wants to probe. In the
specific case of {\lco} and {\srcuocl} that we are considering, the easy
axis is along $b$. Thus, the only nonvanishing matrix elements for
$\chi^x\equiv\chi^a$, which corresponds to the in-plane or $a$ (DM)
mode, are those that mix the spin easy-axis, $b$, with $a$, while for
$\chi^y\equiv\chi^c$, which corresponds to the out-of-plane or $c$
(XY) mode, are those that mix $b$ and $c$. This information allows us 
to establish: (i) to which magnetic-group co-representation the one-magnon 
modes belong, and (ii) the precise scattering geometries needed 
to observe the associated one-magnon Raman modes. For example, it 
tells us that, for either {\srcuocl} or {\lco}, the out-of-plane 
mode (also referred to as XY mode\cite{Gozar,ESR}) belongs to the 
$DB_g$ co-representation of the magnetic point group 
$\underline{m}m\underline{m}$
\be
\lb{chic}
\chi^c\equiv \chi^{XY}=
  \left( \begin{array}{ccc}
     0 & 0 & 0 \\
     0 & 0 & \rmi F \\
     0 & \rmi H & 0
         \end{array} \right),{\;}
\ee
with $C=G=0$. On the other hand, the in-plane mode (also referred to as DM
mode for {\lco}\cite{Gozar,Marcello-Lara}), belongs to the $DA_g$
co-representation of the magnetic point group $\underline{m}m\underline{m}$
\be
\lb{chia}
\chi^a \equiv \chi^{DM}=
  \left( \begin{array}{ccc}
     0 & \rmi B & 0 \\
     \rmi D & 0 & 0 \\
     0 & 0 & 0
         \end{array} \right),{\;}\quad
\ee
with $A=E=I=0$. The precise numerical evaluation of the remaining elements
$B,D,F,H$ requires a more detailed microscopic calculation of the
electric-dipole induced transitions and spin-orbit coupling in second order
perturbation theory, for each system, and this is beyond the scope of this
article. 


\subsection{One-magnon Raman response}

According to the electric-dipole Hamiltonian (\ref{H-ED}), one-magnon Raman
scattering probes the long-wavelength spin excitations $M_i$ around the
staggered spin configuration realized in the antiferromagnetic state. These
correspond to the $({\bf k}=0)$ value of the usual spin-wave dispersions,
where ${\bf k}$ is measured with respect to the ${\bf Q}=(\pi,\pi)$ vector of
the antiferromagentic ordering. In an isotropic Heisenberg antiferromagnet
the spin modes are soft at long wavelength, $\omega_\bk\simeq c k$, but for
anisotropic spin-spin interactions a gap appears at $\bk=0$,
$\omega_\bk=\sqrt{\omega_{a,c}^2+c^2\bk^2}$, where $\omega_{a,c}$ are
related to the anisotropy parameters ($\alpha_a,\alpha_c$ in Eq.\
\pref{conv}, or to $D_+,\Gamma$ in Eq.\ \pref{Hamiltonian}). Starting from
the Hamiltonian \pref{H-ED}, the Raman response can be obtained using
Fermi's golden rule and, for Stokes scattering, we have\cite{Marcello-Lara}
\be 
\lb{Raman-intensity} 
{\cal I}(\omega)=[n_B(\omega)+1]\left\{|\Pi_a|^2 {\cal
A}_{a}({\bf 0},\omega)+  |\Pi_c|^2 {\cal A}_{c}({\bf
0},\omega)\right\}, 
\ee
where ${\cal I}(\omega)$ is the Raman intensity,
$n_B(\omega)=(e^{\beta\omega}-1)^{-1}$ is the Bose function and ${\cal
A}_{a,c}({\bf 0},\omega)$ is the spectral function of the ${\bf k}=0$
transverse $a/c$ components of the antiferromagnetic (staggered) order
parameter. The projectors
\be
\Pi_{a,c}={\bf E}_S^T\chi^{a,c} {\bf E}_I
\ee
are given in terms of the Raman tensors $\chi^{a,c}$ presented above.  The
properties of ${\cal A}_{a,c}$ have been extensively discussed in the
preceeding paper.\cite{LM1}
In particular, it has been shown that the spectral function
${\cal A}_{a,c}({\bf 0},\omega)$ is peaked at the energy $\omega_{a,c}(H)$ of
the magnon gaps in magnetic field, with an intensity $I_{a,c}$ which also
depends on $H$. For {\lco}, $\omega_a$ corresponds to the Dzyaloshinskii-Moriya
gap, while for {\srcuocl} it corresponds to the in-plane gap, as discussed
in the introduction. For both compounds, $\omega_c$ corresponds to the XY
anisotropy gap.

\section{Experimental Setup}

Single crystals of {\lco} and {\srcuocl}\cite{Lance} were measured in a
backscattering geometry with the incoming photons propagating along the $c$
crystallographic axis. The polarization configurations are denoted by
$({\bf E}_I, {\bf E}_S)$ with ${\bf E}_{I/S}$ representing the direction of
the incoming/scattered electric field. The crystals were mounted in a
continuous flow optical cryostat and the Raman spectra were taken using
about 5mW power and the wavelength $\lambda = 647.1$ nm from a Kr$^{+}$
laser. The measurements in magnetic fields were taken with the cryostat
inserted in a room temperature horizontal bore of a superconducting magnet.
The orthorhombic axes of the {\lco} sample were identified by X-ray
diffraction. The data from {\srcuocl} crystal were taken from a freshly
cleaved $(ab)$ surface but in this case the in-plane axes directions were
not determined.

%
\begin{figure}[htb]
\includegraphics[scale=0.6]{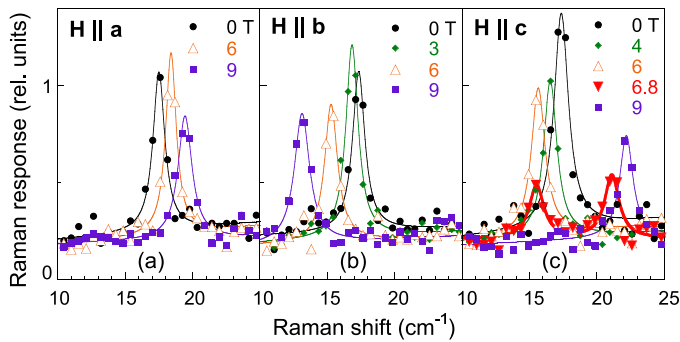}
\caption{(Color online) One-magnon Raman response for {\lco} at $T=10$ K in
  the $(RL)$ circular polarization configuration, where only the
  Dzyaloshinskii-Moriya gap is Raman active. In $(a)$ the gap only hardens,
  in $(b)$ the gap only softens, while in $(c)$ it first softens and then
  hardens above the weak-ferromagnetic transition at $H\simeq 6.6$ T (see
  discussion in the text).}
\label{Fig-Raman-LCO-RL}
\end{figure}
%

We first present the one-magnon Raman response in {\lco} for the $(RL)$
polarization configuration. In this geometry, the electric-field of the
incident light is circularly polarized rotating clockwise, ${\bf
E}_I^R=(\hat{\bf x}_a-\rmi\hat{\bf x}_b)/\sqrt{2}$, while the
electric-field of the scattered light is circularly polarized and rotating
anti-clockwise, ${\bf E}_S^L=(\hat{\bf x}_a+\rmi\hat{\bf x}_b)/\sqrt{2}$.
Here $\hat{\bf x}_a$ and $\hat{\bf x}_b$ are unit vectors along the $a$ and
$b$ directions respectively.  According to our earlier discussion on the
classification of the magnetic excitations in {\lco}, the $(RL)$ in-plane
polarization configuration is the adequate Raman geometry in order to
observe the Dzyaloshinskii-Moriya gap, because it probes directly the 
nonvanishing element $B,D$ of the $\chi^{DM}$ matrix \pref{chia}. The
results are presented in Fig.\ \ref{Fig-Raman-LCO-RL}. In $(a)$ the
magnetic field is applied along the $a$ axis (transverse field). We
observed a monotonic hardening of the gap with increasing magnetic
field. In $(b)$ the magnetic field is applied along the $b$ axis
(longitudinal field). We observed a monotonic softening of the gap
with increasing magnetic field. Finally, in $(c)$ the magnetic field
is applied perpendicular to the {\cuoo} layers (transverse field). We
observed first a softening of the gap with increasing field, a jump at
a field of approximately $6.6$ T, and finally a hardening with
increasing field.

%
\begin{figure}[htb]
\includegraphics[scale=0.6]{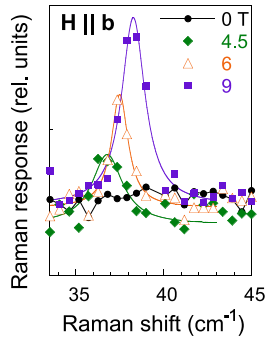}
\caption{(Color online) One-magnon Raman response for {\lco} at $T=10$ K in
  the $(RR)$ circular polarization configuration for a magnetic field
  applied along the orthorhombic $b$ easy-axis, where the
  field-induced mode becomes Raman active (see discussion in the
  text). We observed a monotonic hardening of the field-induced mode
  with increasing applied field.}
\label{Fig-Raman-LCO-RR}
\end{figure}
%

Next we present the one-magnon Raman response in {\lco} for the $(RR)$
polarization configuration. In this geometry, the electric-field of both
the incident and scattered light is circularly polarized rotating
anti-clockwise, ${\bf E}_I^R={\bf E}_S^R=(\hat{\bf x}_a+\rmi\hat{\bf
x}_b)/\sqrt{2}$.  According to the theory described in
Ref. \onlinecite{Marcello-Lara}, the $(RR)$ in-plane polarization
configuration is the adequate Raman geometry in order to observe the
field-induced mode for a magnetic field applied along the orthorhombic $b$
easy-axis, because it probes directly the nonvanishing elements of the {\it
rotated} $\chi^{XY}_\theta$ matrix
\be
\lb{chic-rotated}
\chi^{XY}_\theta=
  \left( \begin{array}{ccc}
     0 & 0 & 0 \\
     0 & \rmi(F+H)\sin\theta\cos\theta & -\rmi H\sin^2\theta+\rmi F\cos^2\theta \\
     0 & -\rmi F\sin^2\theta+\rmi H\cos^2\theta & -\rmi(F+H)\sin\theta\cos\theta
         \end{array} \right),{\;}
\ee
where $\theta$ is the angle of rotation of the spin-quantization basis
within the $(bc)$ plane.\cite{Marcello-Lara} The $(RR)$ polarization
configuration then probes the element $\rmi(F+H)\sin\theta\cos\theta$.  The
results are presented in Fig.\ \ref{Fig-Raman-LCO-RR} and we observed a
monotonic hardening of the gap with increasing magnetic field.  Only the
data points for higher fields are shown in Fig.\ \ref{Fig-Raman-LCO-RR}
because the intensity of the peak rapidly drops for fields lower than $4$
T.

%
\begin{figure}[htb]
\includegraphics[scale=0.6]{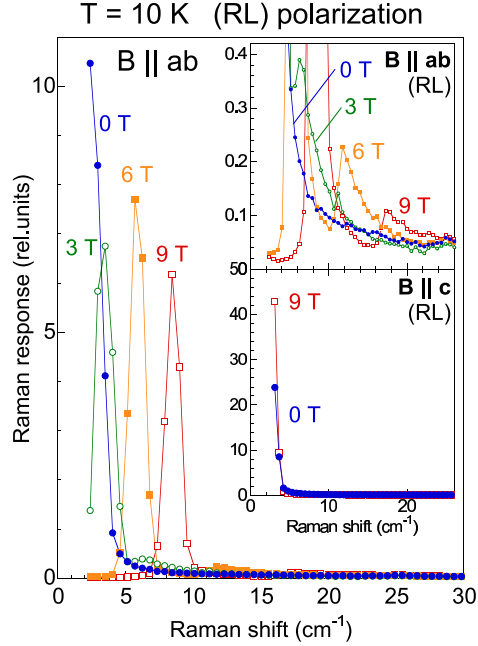}
\caption{(Color online) One-magnon Raman response for {\srcuocl} at $T=10$
  K in the $(RL)$ polarization configuration, where only the in-plane gap
  is Raman active. Main figure: hardening of the in-plane gap with
  increasing in-plane magnetic field. Insets: (top) zoomed immage of the
  main plot, showing the much weaker intensity of the two-magnons
  scattering, starting at approximately twice the magnon gap; (bottom)
  featureless response for perpendicular magnetic field (the out-of-plane
  (XY) mode is not Raman active in the backscattering geometry used here).}
\label{Fig-Raman-SrCuOCl}
\end{figure}
%
 
Finally we present the one-magnon Raman response in {\srcuocl} also for the
$(RL)$ polarization configuration.
The results are presented in Fig.\ \ref{Fig-Raman-SrCuOCl}. First, we
observe that at zero applied field the Raman spectrum is continuously
increasing up to the lowest accessible frequency of 2 cm$^{-1}$. As a
consequence, we can only establish an upper bound of 2 cm$^{-1}$ for the
in-plane magnon gap. This very small value is consistent with the general
expectation that the in-plane gap for {\srcuocl} has a purely quantum
origin.  When the magnetic field is in the plane (main panel) one observes
a hardening of the one-magnon peak, which allows us to clearly identify the
magnon gap as the field increases, while no changes are observed for a
magnetic field parallel to $c$ (bottom inset). Observe also (top inset)
that the signal corresponding to the two-magnon continuum (which starts at
the edge of approximately twice the magnon gap) has much lower intensity
with respect to the one-magnon peak.

\section{Magnetic spectrum in $\mbox{La$_2$CuO$_4$}$}

\subsection{$\bH$ parallel to $a$}

\begin{figure}[htb]
\includegraphics[angle=-90,scale=0.4]{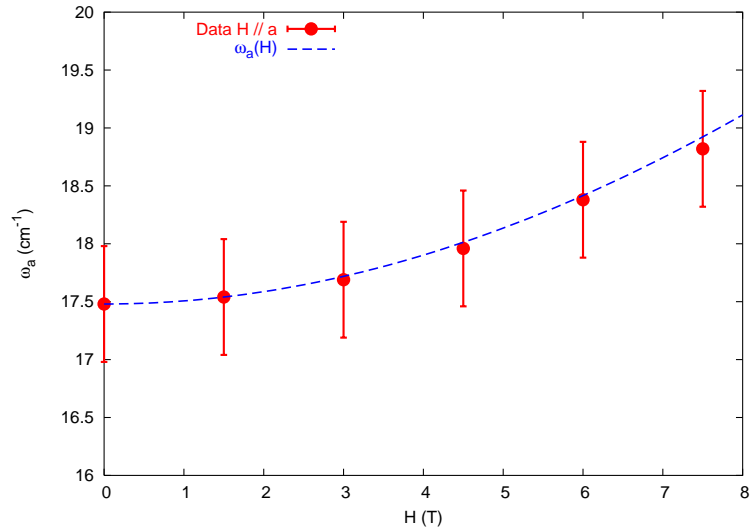}
\caption{(Color online) Comparison between the experimental data and
  theoretical predictions for the field dependence of the DM gap for
  $\bH\parallel a$.  The closed circles are the experimental data. The line
  is the best fit obtained using Eq.\ \pref{gapa}, with $\gamma_a^F=0.93$ (cm
  T)$^{-2}$.}
\label{Fig-gaps-aaxis}
\end{figure}

For a field applied parallel to the $a$ orthorhombic axis, we
obtain the conventional field dependence of the magnon gaps in a
transverse field:\cite{LM1} the hardening of the mode in the field
direction, while the second mode remains unchanged
\be
\lb{hplla}
\omega_a(H)=\sqrt{m^2_a+(g_s^a\mu_B H)^2}, \quad \omega_c(H)=m_c.
\ee
Here we indicate with $m_a, m_c$ the gaps of the magnon modes at zero
field, $\omega_a(0)=m_a$ and $\omega_c(0)=m_c$.  Moreover, with respect to the
preceding article,\cite{LM1} we restored the physical units for the
magnetic field, introducing the Bohr magneton $\mu_B=0.4668$
cm$^{-1}$T$^{-1}$ and the gyromagnetic ratio $g_s^a$ for the field along
$a$. We then obtain for the $\omega_a$ mode the relation
\be \lb{gapa} \omega_a^2=m_a^2+\gamma_a H^2, \quad \gamma_a=(g_s^a \m_B)^2. \ee
As one can see in Fig.\ \ref{Fig-gaps-aaxis}, the data for $\omega_a(\bH\pll
a)$ follow perfectly the previous equation, with a coefficient
$\gamma_a^F=0.93$ (cm T)$^{-2}$ estimated by fitting the data with Eq.\
\pref{gapa}. This value allows us to estimate the gyromagnetic ratio as:
\be
\lb{gsa}
g_s^a=\sqrt{\gamma_a^F}/\m_B=2.0,
\ee
in good agreement with the estimate given usually in the literature.

%
%

\subsection{$\bH$ parallel to $c$}

\begin{figure}[htb]
\includegraphics[scale=0.4]{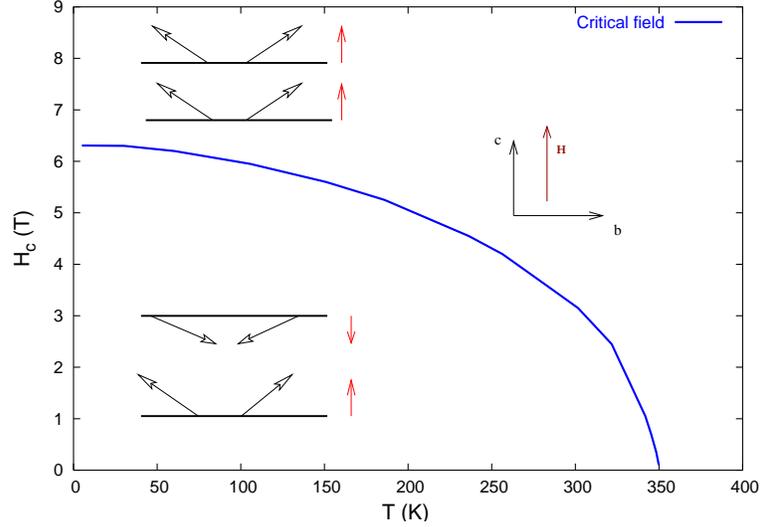}
\caption{(Color online) Phase diagram of the weak-ferromagnetic transition
  and spin configuration for $H\parallel c$. The arrows with open tip
  represent the spins in each layer, the small ones on the right of the
  layers the uniform spin components. At low temperature and zero or low
  field the uniform components of the spins in each layers are ordered
  antiferromagnetically in the $c$ direction. Above the critical field
  \pref{hcrit} the uniform components order ferromagnetically in the $c$
  direction. As the temperature increases the uniform components decrease
  follwing the decrease of the AF spin component along $b$, and as a
  consequence also the critical field for the WF transition decreases. The
  phase diagram has been calculated in Ref.\ [\onlinecite{LM1}] (see
  Fig. 10) by using parameter values appropriate for {\lco}. }
\label{Fig-WF-transition}
\end{figure}

When the field is applied along the $c$ direction one observes a spin-flop
of the ferromagnetic spin components along $c$, which are ordered
antiferromagnetically in neighboring planes at low
field\cite{LM1,Thio90,Papanicolaou}. This weak ferromagnetic
(WF) transition has been indeed measured in
Ref. [\onlinecite{Magnetoresistence}] for the $x=0.01$ doped compound,
occuring at a temperature-dependent critical field of about 4 T at $T=50$
K. Since the magnitude of the ferromagnetic spin components along $c$
decreases with the temperature proportionally to the AF order parameter
$\sigma_0$, also the critical field $H_c$ for the WF transition decreases
with temperature. By properly taking into account the effect of quantum and
thermal fluctuations on the order parameter one can determine the $H_c(T)$
curve reported in Fig.\ \ref{Fig-WF-transition},\cite{LM1} where we also
sketch the spin configuration above and below the transition. However, as
it has been discussed in Ref. \onlinecite{LM1}, a first estimate for the
critical field at low temperature and for the field dependence of the
magnon gaps can be obtained by neglecting the renormalization of 
$\sigma_0$. Using the notation
of Ref. \onlinecite{LM1}, we will decompose the spin at site $i$ in its
staggered $\bn$ and uniform $\bL$ component, so that $\bS_i/S=e^{i\bQ \cdot
{\bf r}_i}\bn_i+a\bL$. In the AF ordered state $\langle \bn \rangle
=\sigma_0$, where in general $\sigma_0$ is corrected both by quantum and
thermal corrections. At low temperatures, and neglecting quantum
fluctuations $(\sigma_0=1)$, one obtains that the critical field is
\be
\lb{hcrit}
H_c=\frac{2\eta }{g_s^c\mu_B D_+},
\ee
where $\eta=2JJ_\perp$ is the energy scale associated to the interlayer
coupling $J_\perp$, and $D_+$ is the modulus of the DM vector, which
controls also the $\bH=0$ value of the DM gap, $D_+\equiv m_a$\cite{LM1}.
%
%
%
%
According to the analysis of Ref. \onlinecite{LM1}, the magnon gaps 
evolve, below $H_c$, as
\bea
\omega^2_a&=&m_a^2 +2\eta-\sqrt{4\eta^2+\left(g_s^c\mu_BHD_+ \right)^2}, \nn\\
\lb{h<hc}
\omega^2_c&=&m_c^2 +(g_s^c\mu_B H)^2+2\eta-\sqrt{4\eta^2+\left( g_s^c\mu_B 
H D_+\right)^2}, \quad H<H_c,
\eea
while above the WF transition they are given by
\bea
\omega^2_a&=&m_a^2 + |g_s^c\mu_B HD_+|,\nn\\
\lb{h>hc}
\omega^2_c&=&m_c^2 +(g_s^c\mu_BH)^2+| g_s^c\mu_B HD_+|, \quad H>H_c.
\eea
Using this set of equations we recognize that the parameter $\eta$
can be determined from the jump of the gap at $H_c$, since
$$
\omega^2_a(H_c^-)=m_a^2-2\eta(\sqrt{2}-1), \quad
\omega^2_a(H_c^+)=m_a^2+2\eta.
$$
We can then estimate
\be
\lb{eta}
2\eta=\frac{\omega^2_a(H_c^+)-\omega^2_{a}(H_c^-)}{\sqrt{2}}=143.25 \, {\mathrm
  cm}^{-2},
\ee
from which it follows also that
\be
\eta=2JJ_\perp \Ra J_\perp/J\sim 3.2 \times 10^{-5},
\ee
where we used $J=130$ meV. We can then recognize that the field evolution
of the $\omega_a$ gap above and below the WF transition is controlled by a
single parameter $\gamma_c$
\bea
\omega^2_a&=&m_a^2 +2\eta-\sqrt{4\eta^2+(\gamma_c H)^2}, \quad H<H_c\nn\\
\lb{gapc}
\omega^2_a&=&m_a^2 + \gamma_c H, \quad H>H_c,
\eea
where $\gamma_c$ is uniquely determined by $2\eta$, $H_c$ and
the gyromagnetic ratio of the $c$ direction $g_s^c$
\be
\lb{gammac}
\gamma_c=g_s^c \m_B D_+ =\frac{2\eta}{H_c}.
\ee
As a consequence, we can extract $g_s^c$ both by fitting the experimental
data with Eqs. \pref{gapc} and obtaining a value $\gamma_c^F$, and by using
directly the last two equalities of Eq.\ \pref{gammac}. In the former case
we obtain $\gamma_c^F=$ 21.83 cm$^{-2}$T$^{-1}$, which corresponds to
$g_s^c=2.7$.  In the latter case instead we can use the value of $2\eta$
obtained from the jump of the gap at $H_c$, the value $D_+=m_a=17.48$
cm$^{-1}$, and the measured value $H_c\simeq 6.6$ T, finding
\be
\lb{gsc}
g_s^c=\frac{2\eta}{\mu_B H_c D_+}=2.65.
\ee
Thus, both estimates give a value which is quite close to the one commonly
quoted in the literature, $g_s^c=2.45$.

%
\begin{figure}[htb]
\includegraphics[angle=-90,scale=0.4]{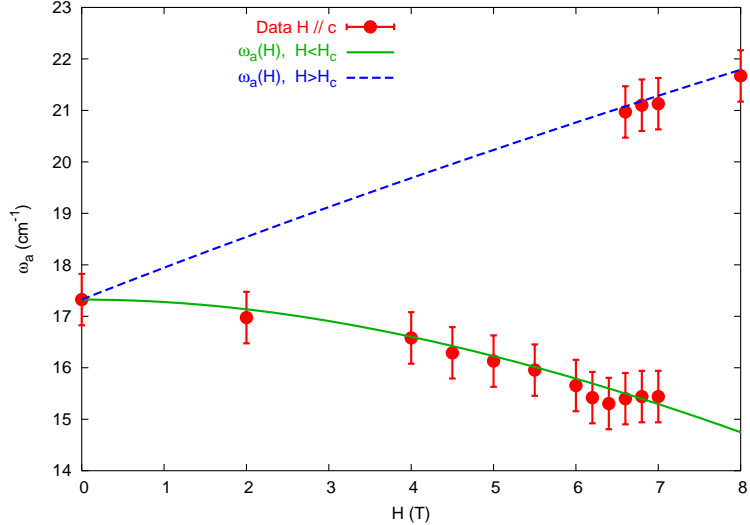}
\caption{(Color online) Comparison between the experimental data and the
  theoretical predictions for the field dependence of the DM gap for
  $\bH\pll c$.  The closed circles are the experimental data. Observe that
  near the WF transition two values of the gap have been measured (see also
  panel (c) of Fig.\ \ref{Fig-Raman-LCO-RL}). Indeed, since the transition
  is first order in a finite sample the transition is observed as a
  coexistence of the two phases (above and below the transition) for a
  range of field values around the thermodynamic critical field. The lines
  are the fit using Eqs. \pref{gapc}, with $\gamma_c^F=21.83$
  cm$^{-2}$T$^{-1}$.}
\label{Fig-gaps-caxis}
\end{figure}
%

Observe that this value of $g_s^c$ is only an upper bound because we did
not include in the set of Eqs. \pref{gapc} and in the definition of the
critical field \pref{hcrit} the quantum and thermal correction of the order
parameter $\sigma_0$, which reduce its value with respect to $\sigma_0=1$
used so far. As it has been explained in Ref. \onlinecite{LM1}, when this
effect is taken into account one must replace $D_+$ by $D_+/\sigma_0$ in
Eqs. \pref{h<hc}-\pref{h>hc}
\bea
\omega^2_a&=&m_a^2 +2\eta-\sqrt{4\eta^2+\left(\frac{g_s^c \mu_B H
    D_+}{\sigma_0}\right)^2}, \nn\\
\lb{h<hcnew}
\omega^2_c&=&m_c^2 +(g_s^c \mu_B H)^2+2\eta-
\sqrt{4\eta^2+\left(\frac{ g_s^c \mu_B H D_+}{\sigma_0}\right)^2}, \quad H<H_c,
\eea
and
\bea
\omega^2_a&=&m_a^2 +\left( \frac{ g_s^c \mu_B H
    D_+}{|\sigma_0|}\right),\nn\\
\lb{h>hcnew}
\omega^2_c&=&m_c^2 +(g_s^c \mu_B H)^2+\left( \frac{g_s^c \mu_B H
    D_+}{|\sigma_0|}\right), \quad H>H_c.
\eea
At the same time the critical field is a function of $\sigma_0$ according to
\be
\lb{hcritnew}
H_c=\frac{4\eta}{g_s^c \mu_B D_+}\frac{\sigma_0}{1+\sigma_0^2}.
\ee
The order parameter $\sigma_0$ is determined at each temperature by computing
the effect of transverse spin-wave fluctuations, which depends on
the magnon gap according to a general equation like
\be
\lb{s0}
\sigma_0^2(H,T)=1-I_\perp(\omega_a(H),\omega_c(H),T),
\ee
where the precise expression for $I_\perp$ is given in Ref.
\onlinecite{LM1}. Note that since the DM interaction introduces an explicit
dependence of the magnon gaps on the order parameter, see Eqs.\
\pref{h<hcnew}-\pref{h>hcnew}, Eq.\ \pref{s0} is a self-consistency
equation for $\sigma_0$.  As far as the previous estimates of $\eta$ and 
$g_s^c$ are concerned, one can see that the solution of the full set 
of Eqs.\ \pref{h<hcnew}-\pref{h>hcnew} \pref{hcritnew} and \pref{s0} 
can slightly modify the values previously obtained in two respects. 
First, the jump of the gap at the critical field will depend not only 
on $\eta$ but also on $\sigma_0$. Second, the analogous of Eq.\ \pref{gsc} 
will be
\be
g_s^c=\frac{4\eta}{\mu_B H_c D_+}\frac{\sigma_0}{1+\sigma_0^2}.
\ee
Observe that both these corrections will contribute to a reduction of
$g_s^c$ with respect to the previous estimate. However, since the
experimental determination of $\sigma_0$ in this sample is not available
and the theoretical value depends on the approximations used to derive Eq.\
\pref{s0}, we retain here the estimates given above using $\sigma_0=1$ and
we refer the reader to Ref. \onlinecite{LM1} for a more detailed discussion
of this issue.

We should point out that a somewhat larger estimate for the critical field
$H_c$ has been extracted recently from the neutron scattering measurements
of Ref. \onlinecite{Reehuis}, in a {\lco} sample with almost the same
N\'eel temperature as the one considered here. According to the previous
Eq.\ \pref{hcritnew}, several factors can affect the critical field. Thus,
the larger critical field measured in the sample of
Ref. \onlinecite{Reehuis} can be explained with a larger value of the
staggered order parameter and of the interlayer coupling $\eta$, or also
with a smaller value of the DM vector $D_+$. However, one should remember
that the weak-ferromagnetic transition for $H \parallel c$ is a first-order
transition accompanied by a hysteresis that can be large at low temperature
(see for example Ref. \onlinecite{Magnetoresistence}), and hence the
experimentally determined critical field could be affected by the
hysteresis. As a consequence, further investigation of the parameter values
for this sample can shed more light on this apparent discrepancy between
the Raman and the neutron scattering measurements.

\subsection{$\bH$ parallel to $b$}

Finally, let us discuss the case of $\bH\pll b$, where also the data for
the field-induced mode $\omega_c$ are available for $H\geq 4$ T. Here again
the field dependence of the magnon gaps follows a different behavior for
small and large field. As it has been explained in Ref. \onlinecite{LM1},
due to the DM interaction a field along $b$ generates an effective
staggered field in the $c$ direction, giving rise to a rotation of the
staggered order parameter in the $bc$ plane.\cite{Reehuis} As a
consequence, at low field the classical configuration of the staggered AF
order parameter $\bn_m$ in the $m$-th plane is given by
\be
\lb{solb1}
\bn_m=(0,\sigma_b,(-1)^m\sigma_c).
\ee
where $\sigma_b$ and $\sigma_c$ indicate, respectively, the components of
the order parameter along $b$ and $c$ direction. Observe that an
oscillating staggered $\sigma_c$ component implies that in the spin
decomposition $\bS_i/S=e^{i\bQ\cdot {\bf r}_i}\bn_i+\bL_i$ given above
actually the $S^c$ components coming from $n^c$ are ordered {\em
ferromagnetically} in neighboring planes, to allow for the (average)
uniform spin components $\bL$ induced by the DM vector to align along the
applied field.\cite{LM1} 

\begin{figure}[htb]
\includegraphics[scale=0.4]{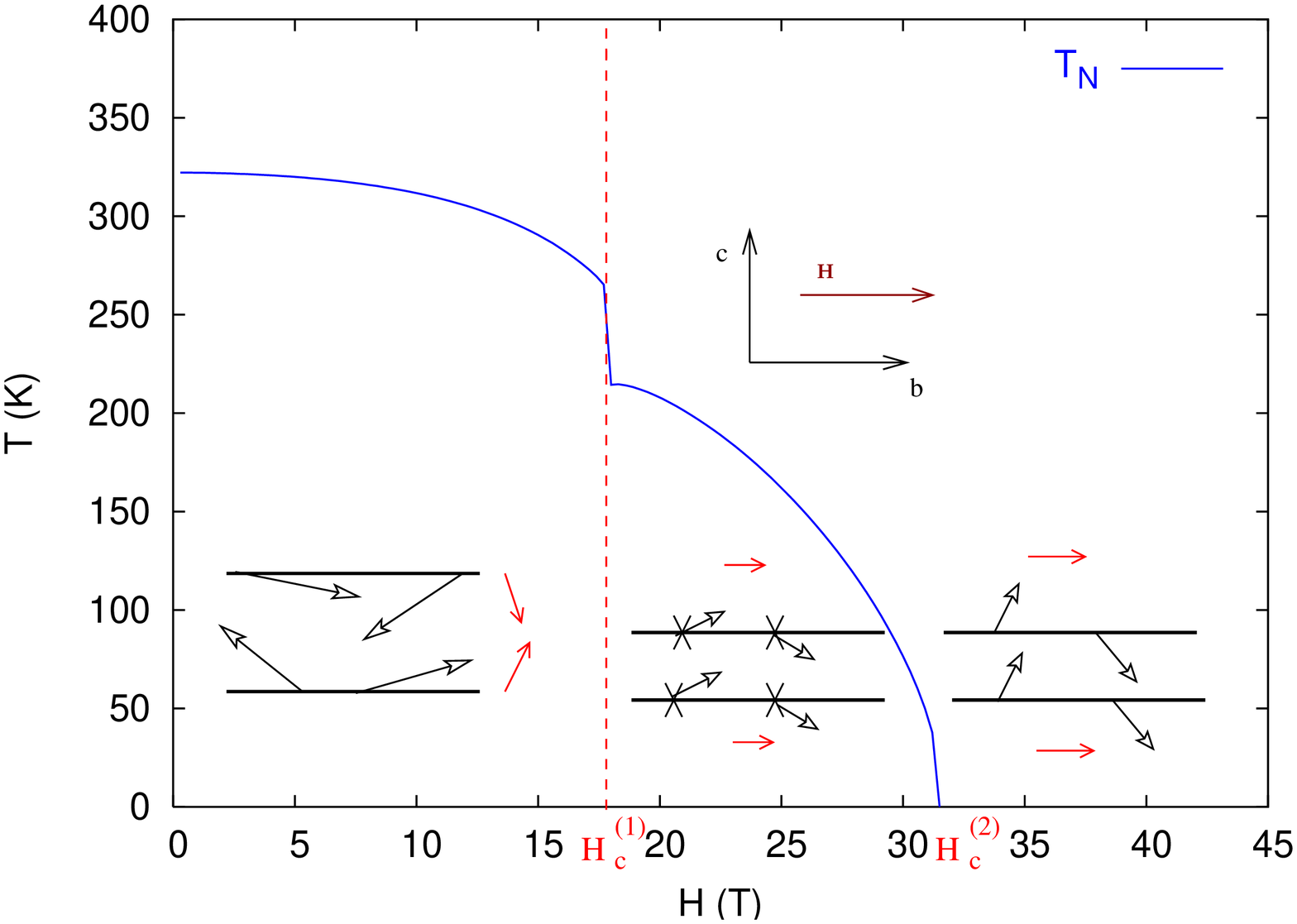}
\caption{(Color online) Phase diagram and evolution of the spin
configuration for $H\parallel b$. For the spin components we use the same
notation as in Fig.\ \ref{Fig-WF-transition} above.  At $H<H_c^{(1)}$ both
the spins and the uniform components rotate in the $bc$ plane, so that the
net uniform magnetization in along the $b$ axis. In this region $T_N(H)$ is
the temperature at which the in-plane AF component along $b$ vanishes. At
$H_c^{(1)}<H<H_c^{(2)}$ the AF spin components allign along $a$ which is
perpendicular to the plane of the figure, so that these spin components,
which vanish at the temperature $T_N(H)$, are represented by the small
crosses. At the same time, the uniform components align ferromagnetically
in neighboring planes.  At $H>H_c^{(2)}$ only the $c$ components survive.}
\label{Fig-Rotation}
\end{figure}

As the field strenght increases, the spins perform
a two-step spin-flop transition:\cite{LM1,Thio90,Papanicolaou} at
$H=H_c^{(1)}$ the in-plane AF components allign along the $a$ direction,
while at $H=H_c^{(2)}>H_c^{(1)}$ the $a$ component vanishes and only the
staggered spin components in the $c$ direction are left. In Fig.\
\ref{Fig-Rotation} we shown the $(H-T)$ phase diagram which has been
evaluated in Ref.\ [\onlinecite{LM1}] by means of a saddle-point
approximation for the transverse spin fluctuations. The temperature $T_N$
is the one where the in-plane spin component (along $b$ or $a$) vanishes,
leaving a non-zero $\sigma_c$ component. Observe that the jump of $T_N(H)$
at $H_c^{(1)}$ is an artifact of the approximation: indeed,
the critical field for
this transition is controlled by the energy scale where the $\omega_a$ gap
vanishes (see Eq.\ \pref{hpllb} below), i.e.
\be
\lb{hc1}
g_s^b\mu_B H_c^{(1)}=m_a=D_+,
\ee
which corresponds, in our case, to $H_c^{(1)}=18$ T. Since within the
saddle-point approximation $m_a$ is temperature independent, at this level
one cannot recover the temperature variation of $H_c^{(1)}$ which would
eliminate the anomalous jump of $T_N(H)$ reported in Fig.\
\ref{Fig-Rotation}.  Since the maximum field used in the present experiments is
$H=11$ T, this spin flop transition is out of the range accessible in our
measurements. It is perhaps worth pointing out that in a recent
work the evolution of the magnetic signal for $H\pll b$ has been measured
by means of neutron scattering up to fields of 14 T.\cite{Reehuis} By
assuming a power-law decay of the intensity of the $(100)$ neutron peak -
corresponding to magnetic order along the $b$ direction - the authors could
estimate for the above critical field a slightly higher value of $22$
T. Considering that a theoretical prediction of the neutron-peak intensity
as a function of magnetic field is still lacking, this estimate seems in
good agreement with the value \pref{hc1} given above.

%
%

\begin{figure}[htb]
\includegraphics[angle=-90,scale=0.4]{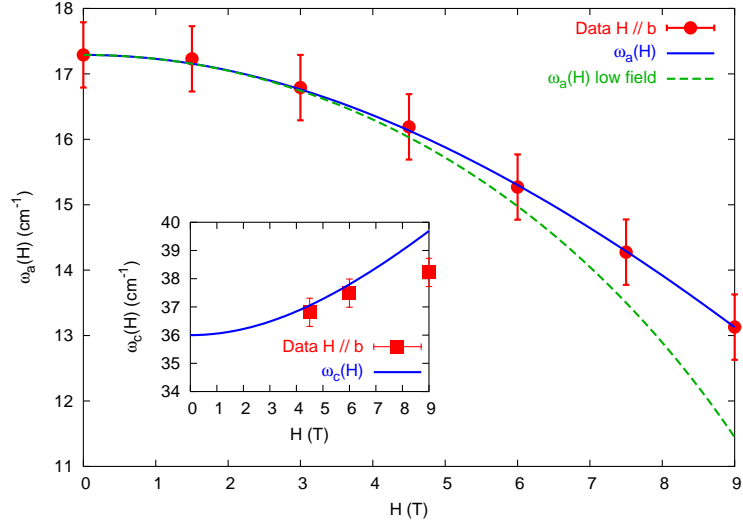}
\caption{(Color online) Comparison between the experimental data and the
  theoretical predictions for the field dependence of the $a$ mode for
  $\bH\parallel b$. The blue solid line is the fit done using the first of
  Eqs.\ \pref{hpllb} with $g_s$ and $m_c$ as fitting parameters, giving
  $g_s^b=2.08$ and $m_c=36$ cm$^{-1}$. The green dashed line is the
  approximated expression \pref{mx} evaluated with the same parameters
  values, and it is only valid at low field. Inset: field dependence of the
  $c$ mode observed at fields larger than 4 T. The solid line is the curve
  corresponding to the second of Eqs.\ \pref{hpllb}.}
\label{Fig-gaps-baxis}
\end{figure}

As far as the magnon-gap evolution with magnetic field is concerned, at
field below $H_c^{(1)}$ it is the usual one for longitudinal
fields\cite{LM1}
\bea
{\omega_a^2}=\frac{m^2_a+m_c^2}{2}+(g_s^b\mu_B H)^2-
\sqrt{\left(\frac{m^2_a-m_c^2}{2}\right)^2+4(g_s^b\mu_BH)^2
  \left(\frac{m^2_c+m_a^2}{2}\right)}, \nn\\
\lb{hpllb}
{\omega_c^2}=\frac{m^2_a+m_c^2}{2}+(g_s^b\mu_BH)^2+
\sqrt{\left(\frac{m^2_a-m_c^2}{2}\right)^2+4(g_s^b\mu_BH)^2
  \left(\frac{m^2_c+m_a^2}{2}\right)}.
\eea
At small field these expressions can be approximated as
\bea
\lb{mx}
\omega^2_a=m_a^2-\gamma_a H^2, \quad \gamma_a=(g_s^b\mu_B)^2\D m^a, \quad \D m^a=\left[
2\frac{m^2_c+m^2_a}{m^2_c-m^2_a}-1\right]\\
\omega^2_c=m_c^2+\gamma_c H^2, \quad \gamma_c=(g_s^b\mu_B)^2\D m^c, \quad \D m^c=\left[
2\frac{m^2_c+m^2_a}{m^2_c-m^2_a}+1\right]\nn
\eea
Using Eq.\ \pref{hpllb} and the zero-field value of the $a$ mode we obtain
again an excellent agreement with the experimental points, as one can see
in Fig.\ \ref{Fig-gaps-baxis}. For the sake of completeness we also report
in Fig.\ \ref{Fig-gaps-baxis} the approximate expression \pref{mx}, which
is indeed valid until the field $H_l\simeq 4$ T. From the fit of the $\omega_a$
data using Eqs.\ \pref{hpllb} we obtain both $m_c=\omega_c(H=0)$ and the
gyromagnetic ratio. The results are
\be
\lb{est}
m_c=36 \, \mathrm{cm}^{-1}, \quad g_s^b=2.08,
\ee
which are again in excellent agreement with the values reported in the
literature. 

In the inset of Fig.\ \ref{Fig-gaps-baxis} we show also the
field dependence of the field-induced $\omega_c$ (XY) mode according to
Eq.\ \pref{hpllb}, using the values \pref{est} extracted from the
fitting of the $\omega_a$ mode. As one can see, only the data point at $9$ T
is out of the range of the thoeretical curve. The appearence of the 
field-induced $\omega_c$ mode for the in-plane Raman scattering in the 
$(RR)$ polarization configuration is now well understood as being a 
consequence of a continuous rotation of the magnetization axis when 
$\bH \pll b$.\cite{Marcello-Lara} 

%
%
%

\subsection{Peak intensity as a function of magnetic field}

A complete microscopic theory of the Raman scattering which allows one to
compute the exact shape and the absolute intensity of the Raman peaks is
out of the scope of the present paper. Nonetheless, when data taken at
different magnetic fields are compared, one could expect that the main
dependence of the relative peak intensity measured by Raman can be at least
qualitatively described by the theory developed in Ref.
\onlinecite{LM1}. Indeed, the spectral function ${\cal A}_{a,c}$ contains
itself an intrinsic dependence of the peak intensity on the magnetic field,
which enters in the Raman response trough the relation
\pref{Raman-intensity}. Thus, we can evaluate Eq.\ \pref{Raman-intensity}
using the theoretical prediction for ${\cal A}_{a,c}$ and compare it with
the experimental data. As we shall see, even though our analysis does not
include the spin damping processes, the overall agreement between the
theoretical predictions and the experimental data is fairly good.

As it has been discussed in Ref. \onlinecite{LM1}, the spectral
function of each mode is defined from the Green's function for the
corresponding fluctuations. In the absence of magnetic field the Green's
function for the transverse mode is a diagonal matrix
\be
\lb{gm}
\hat G^{-1}=
\frac{1}{gc}\left( \begin{array}{cc}
\omega_n^2+\varepsilon_a^2(\bk) &  0 \\
 0       & \omega_n^2+\varepsilon_c^2(\bk)
\end{array} \right),
\ee
where the magnon gap is by definition
$\omega_{a,c}=\varepsilon_{a,c}(\bk=0)$. In this case the Green's function
matrix is also diagonal and we simply obtain
\be
\lb{ca-st}
\cA_a(\omega>0)=-\frac{1}{\pi}Im G_a(i\omega_n\ra \omega+i0^+)=
\frac{1}{2\omega_a}\d(\omega-\omega_a),
\ee
and analogously for $\cA_c$.  Thus, from Eq.\ \pref{ca-st} we can deduce that
the peak intensity $I_a$ evolves as $1/\omega_a$, i.e. it is larger for smaller
gap values. 

When a magnetic field is applied, two different cases must be considered:
(i) if the matrix $\hat G$ is still diagonal the structure \pref{ca-st} of
the spectral function is preserved, and both the peak position and its
intensity evolve as $\omega_{a,c}(H)$, and $I_a=1/\omega_{a,c}(H)$,
respectively; (ii) if off-diagonal terms proportional to the magnetic field
appear in Eq.\ \pref{gm} the Green's function of the transverse
fluctuations has a non-diagonal structure which leads to the appearance of
several magnon gaps in the response of each single mode. For example, in
the case of longitudinal field one has for the spectral function of the $a$
mode a structure like\cite{LM1}
\be
\lb{ca}
\cA_a(\omega>0)=
\left[\frac{Z_a}{2\omega_a}\d(\omega-\omega_a)+\frac{Z_c}{2\omega_c}\d(\omega-\omega_c)\right].
\ee
However, one finds in general that $Z_a/\omega_a\gg Z_c/\omega_c$, so that
essentially a single peak at the $\omega_a$ energy is observed in the
measurements, but the spectral weight of this peak is $I_a=Z_a/\omega_a$
and not just $1/\omega_a$ as in Eq.\ \pref{ca-st}.  Indeed, the factor
$Z_a$ in Eq.\ \pref{ca} leads in general to an additional field dependence
of the intensity on the magnetic field.

\begin{figure}[htb]
\includegraphics[angle=-90,scale=0.4]{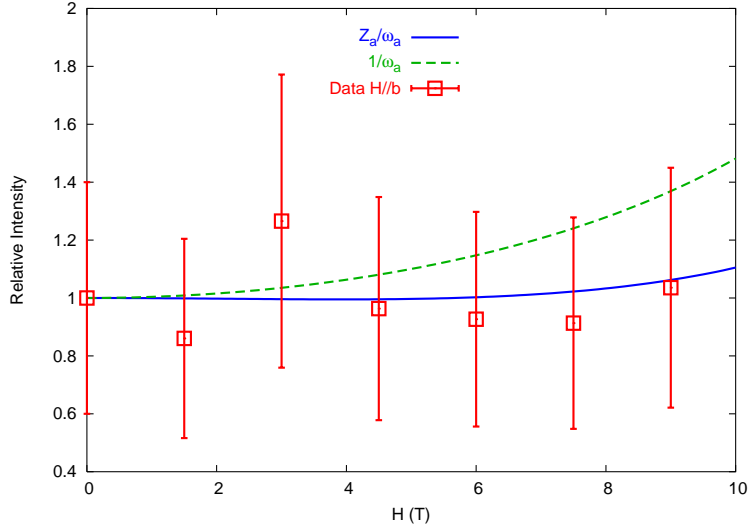}
\caption{(Color online) Field dependence of the normalized intensity
  $I_a(H)/I_a$ of the $a$ peak for a longitudinal field. We estimated the
  relative error on the measured peak intensity around 20 $\%$. Solid line:
  field dependence of the peak intensity according to Eqs.\ \pref{ca} and
  \pref{za}. Observe that by neglecting the contribution of the $Z_a$
  factor from Eq.\ \pref{za} one would obtain $I_a=1/\omega_a$, corresponding
  to the dashed line, which shows an increasing of the relative intensity
  not observed in the experiments.}
\label{Fig-int-baxis}
\end{figure}

For an ordinary easy-axis AF Eq.\ \pref{ca-st} is valid when a transverse
field is applied, so that an hardening of the gap in the field direction
should be also accompanied by a softening of the peak intensity. For a
longitudinal field one finds instead the spectral function \pref{ca} given
above. According to the calculations of Ref. \onlinecite{LM1}, $Z_a$ is
given by
\be
\lb{za}
Z_a=\frac{-\omega^2_{a}+m^2_c-H^2}{\omega_c^2-\omega_a^2},
\ee
and it is a decreasing function of the magnetic field. As a consequence,
the overall factor $Z_a/\omega_a$ in Eq.\ \pref{ca} is increasing much more
slowly than the $\sim 1/\omega_a$ behavior that one could expect for a
transverse gap, as it is shown in Fig.\ \ref{Fig-int-baxis}. Here,
according to Eq.\ \pref{Raman-intensity}, we included also the bose factor
$[n_B(\omega)+1]$, whose contribution to the overall dependence of the peak
intensity is however very small.
%
%
%
%

\begin{figure}[htb]
\includegraphics[angle=-90,scale=0.4]{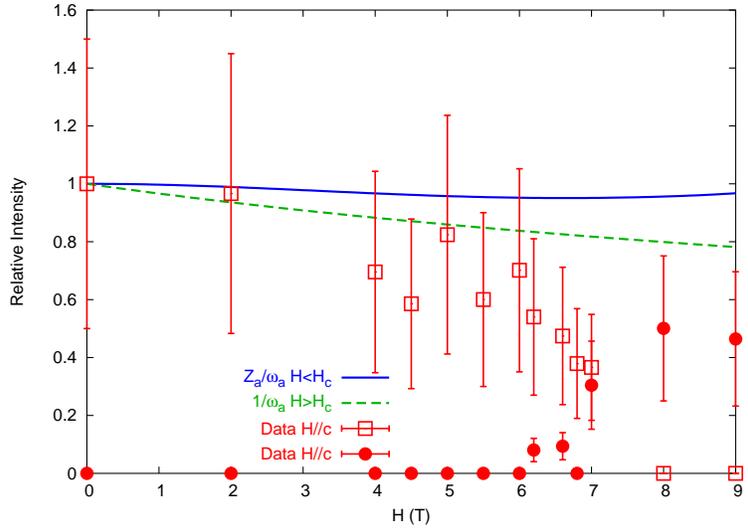}
\caption{(Color online) Field dependence of the intensity of the $a$ peak
  for a transverse field in the $c$ direction. Observe that the spin-flop
  WF transition is first order, so that by increasing the magnetic field one
  first access a crossover regime where the Raman intensity is split
  between the two peaks corresponding to the equilibrium gap values at
  $H<H_c$ and $H>H_c$. Thus, only the data at $H>7$ T should be compared
  with the dashed curve, corresponding to the state above the spin flop.}
\label{Fig-int-caxis}
\end{figure}

In the case of $\bH\pll c$ in an ordinary easy-axis AF one would expect the
peak for the $a$ to be unchanged, as indeed observed in {\srcuocl}. However,
for {\lco} the presence of the DM interaction does modify both the peak
position and its intensity as a function of magnetic field. Below the
critical field for the spin-flop transition the spectral function of the
$a$ mode is\cite{LM1}
\be
\cA_{a}(\omega>0)\approx \frac{Z_a}{\omega_a}\d(\omega-\omega_a),
\ee
where
\be
\lb{zm}
Z_{a}=
\frac{2\eta+\sqrt{4\eta^2 +(HD_+/\sigma_0)^2}}
{2 \sqrt{4\eta^2 +(HD_+/\sigma_0)^2}}.
\ee
Once again, while the $1/\omega_a(H)$ increases with magnetic field the
$Z_a$ factor decreases, giving rise to an almost constant spectral weight
below the spin-flop transition, see Fig.\ \ref{Fig-int-caxis}. Here we used
the $\eta$ value extracted from the fit with $\sigma_0=1$. Above the
spin-flop transition the matrix of the transverse fluctuations is again
diagonal and the standard $I_a=1/\omega_a$ spectral weight is expected,
with the $\omega_a(H)$ dependence given by Eq.\ \pref{h>hcnew}.

%
%

\section{Magnetic Spectrum in $\mbox{Sr$_2$CuO$_2$Cl$_2$}$}

The case of {\srcuocl} is rather simple. This system is very well described
by the Hamiltonian \pref{conv} with a large XY anisotropy $\alpha_c\sim
10^{-4}J$.  Since the system is tetragonal, one does not have in principle
any intrinsic in-plane anisotropy, i.e. one would expect $\alpha_a=0$ in this
case. Nonetheless, it has been suggested in Ref. \onlinecite{Yildirim} that a
very small in-plane gap can still exist as due to purely quantum
effects. Recent ESR measurements seem to confirm this prediction and give
an estimate of the in-plane gap as $m_a=0.048\sim 0.05$ meV.\cite{ESR} As
we can see from Fig. \ref{Fig-Raman-SrCuOCl} there might be a peak at low
energies (below 2 cm$^{-1}$ for $H=0$ T), which can well correspond to the
same in-plane gap as observed by ESR. As far as the field dependence of the
gaps is concerned, the results obtained in the Sec. III-IV of the preceding
article for a conventional easy-axis AF still apply.\cite{LM1} Moreover,
since no orthormonbic distortion exists in {\srcuocl}, no DM interaction is
present, and no effects due to staggered fields will show up in the field
dependence of the magnon gaps.

\begin{figure}[htb]
\includegraphics[angle=-90,scale=0.4]{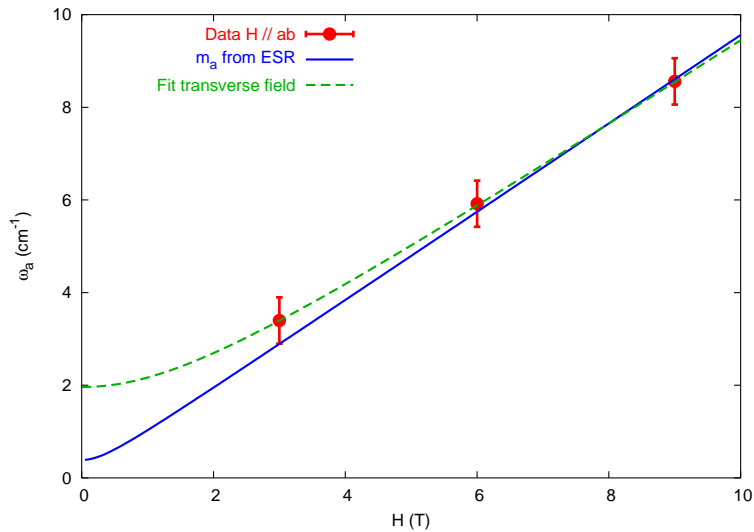}
\caption{(Color online) Field dependence of the in-plane spin-wave gap
  measured by Raman in {\srcuocl}. Solid line: field dependence
  of the $a$ gap obtained using the value $m_a=0.4$ cm$^{-1}$ and
  $g_s=2.05$ measured by ESR and the theoretical prediction for a
  transverse field. Dashed line: best fit on the data
  point using Eq.\ \pref{In-Plane-Gap-SCOC} with $m_a$ and $g_s$ as
  fitting parameters.}
\label{Fig-gap-SCOC}
\end{figure}

In Fig. \ref{Fig-gap-SCOC} we compare the extracted dispersion of the
in-plane gap in {\srcuocl} for an applied magnetic field parallel to
the {\cuoo} layers, with the predictions for a transverse magnetic
field from the theory presented in the preceding article\cite{LM1} 
%
\be
{\omega_a}(H)=\sqrt{m_a^2+(g_s\mu_B H)^2}.
\label{In-Plane-Gap-SCOC}
\ee
As we can see, by using the value of the gap given by the ESR measurements,
$m_a=0.048\sim 0.05$ meV,\cite{ESR} and $g_s=2.05$ the agreement is already
quite good. On the other hand, we can use the Raman data alone to give an
independent estimate of both $m_a$ and $g_s$, in the same spirit of the
previous Sections. This is shown in Fig.\ \ref{Fig-gap-SCOC}, and the best
fit is for $g_s=1.98$ and $m_a=1.96$ cm$^{-1}$ ($m_a=0.24$ meV). Observe
that this estimate of $m_a$ is larger than the one measured by ESR, but it
is still consistent with the fact that at $H=0$ no gap was observed in the
Raman spectra at frequency larger than 2 cm$^{-1}$, which is the lower
bound of the accessible frequency range as shown in Fig.\
\ref{Fig-Raman-SrCuOCl}.


%
%

Finally, we should emphasize that no field-induced modes were observed
in {\srcuocl} neither for perpendicular nor in-plane magnetic
fields. This results from the absence of the Dzyaloshinskii-Moriya
interaction in {\srcuocl}. In fact, it is exactly the
DM interaction in {\lco} that causes the rotation
of the spin-quantization basis and modifies the Raman selection rules
allowing for the appearance of a field-induced mode when the field is along
the easy axis.\cite{Marcello-Lara} Moreover, the DM
interaction is responsible for the in-plane gap evolution for a field along
$c$, which is instead unexpected in a conventional easy-axis
antiferromagnet. 

\section{Conclusions}

We have measured, discussed, and compared the one-magnon Raman spectrum of
{\srcuocl} and {\lco}. We have seen that, for the case of {\srcuocl}, which
is a conventional easy-axis antiferromagnet, there is an in-plane magnon
mode at very low energies in the $DA_g$ channel that is accessible in the
$(RL)$ polarization configuration. No out-of-plane nor field-induced modes
were observed in the geometries used, in agreement with the general
expectation for a conventional easy-axis antiferromagnet.  For the case of
{\lco}, the magnetic field evolution of the in-plane gap measured in the
$(RL)$ polarization configuration is made rather nontrivial due to the
presence of the DM interaction. When the field is along $c$ the DM gap
first softens, jumps discontinuously at the critical field for the WF
transition and then hardens. At the same time, when a longitudinal field is
applied the $a$ mode softens and the (field-induced) $c$ mode appears in
the $(RR)$ polarization, as a consequence of a rotation of the spin
quantization basis.\cite{Marcello-Lara}. The magnetic-field dependence of
the one-magnon Raman energies were found to agree remarkably well with the
theoretical predictions of part I of this work,\cite{LM1} allowing us to
extract from the Raman spectra the values of the various components of the
gyromagnetic tensor and of the interlayer coupling. Moreover the analysis
of the field-evolution of the Raman-peak intensity, which also shows a
general good agreement with the experiments, demonstrated that the
long-wavelength analysis of Ref. \onlinecite{LM1} contains new useful
informations with respect to the standard spin-wave calculations of the
magnon gaps existing in the literature.

\section{Acknowledgements}
The authors would like to acknowledge helpful discussions with
R.~Gooding and B.~Keimer.


\begin{thebibliography}{20}
\bibitem{Discussion-NiF2}
It is perhaps worth pointing out that exactly the same situation
occurs for the case of NiF$_2$. NiF$_2$ has a rutile crystal structure
and in the paramagnetic phase the point group is the tetragonal
$I4/mmm$ group. Below $T_N$, however, the magnetic point group is the
$\underline{m}\underline{m}m$ group exactly because the spin easy-axis
is not parallel to the $4$-fold axis of symmetry. For MnF$_2$ on
the other hand, where the easy-axis is in fact parallel to the
$4$-fold axis, the magnetic group is also tetragonal (see Birss' book
\cite{Birss}).
\end{thebibliography}
\end{document}